# Revealing spatially heterogeneous relaxation in a model nanocomposite

Shiwang Cheng,[1] Stephen Mirigian,[2] Jan-Michael Y. Carrillo,[3,4] Vera Bocharova,[1] Bobby G. Sumpter,[3,4] Kenneth S. Schweizer,[2] Alexei P. Sokolov[1,5*]

[1] Chemical Sciences Division, Oak Ridge National Laboratory,
Oak Ridge, Tennessee 37831, USA

[2] Department of Materials Science and Chemistry, Frederick Seitz Materials Research Laboratory, University of Illinois, Urbana, Illinois 61801, USA

[3] Center for Nanophase Materials Sciences, Oak Ridge National Laboratory,
Oak Ridge, Tennessee 37831, United States

[4] Computer Science and Mathematics Division, Oak Ridge National Laboratory,
Oak Ridge, Tennessee 37831, United States

[5] Department of Chemistry, Department of Physics and Astronomy,
University of Tennessee, Knoxville, Tennessee 37996, USA

The detailed nature of spatially heterogeneous dynamics of glycerol-silica nanocomposites is unraveled by combining dielectric spectroscopy with atomistic simulation and statistical mechanical theory. Analysis of the spatial mobility gradient shows no 'glassy' layer, but the $\alpha$-relaxation time near the nanoparticle grows with cooling faster than the relaxation time in the bulk, and is ~20 times longer at low temperatures. The interfacial layer thickness increases from ~1.8 nm at higher temperatures to ~3.5 nm upon cooling to near $T_g$. A real space microscopic description of the mobility gradient is constructed by synergistically combining high temperature atomistic simulation with theory. Our analysis suggests that the interfacial slowing down arises mainly due to an increase of the local cage scale barrier for activated hopping induced by enhanced packing and densification near the nanoparticle surface. The theory is employed to predict how local surface densification can be manipulated to control layer dynamics and shear rigidity over a wide temperature range.

---

[*] Corresponding author.   Email: sokolov@utk.edu

# I. Introduction

Nanocomposite materials are widely used in many advanced applications, including lightweight materials, coatings, membranes, and solar and fuel cells [1-3] due to improved mechanical, thermal, optical, electrical and catalytic properties.[1-7] It is believed that the large interfacial area between nanoparticles and the host matrix plays a central role in the often observed favorable macroscopic property changes which are nucleated by the modification of matrix packing, relaxation and elasticity near the particle surface.[3, 6, 8] Thus, a clear understanding of the spatially heterogeneous structure and dynamics in interfacial layers is crucial for the rational design of nanocomposites with performances on demand. This fundamental problem also arises in the geometrically simpler, but technologically important, context of supported and capped thin films.[9-11]

Since the discovery of different mobility domains in filled elastomers by NMR,[12, 13] additional experiments,[14-26] theories [27-30] and computational simulations [23, 26, 31-34] have been applied to study the interfacial properties in polymer nanocomposites (PNCs). Owing to strong polymer-particle attractions required for miscibility and particle dispersion, a "glassy" or "dead" layer (defined as an $\alpha$-relaxation time > 100 s) ~1-6 nm thick is often suggested.[14-16, 24, 25] However, this interpretation is controversial [20-22, 26] and the answer may be chemically specific. Current experimental limitations and inability to spatially resolve dynamics close to a nanoparticle surface provide significant obstacles in studies of the interfacial mobility gradient. Moreover, computer simulations can probe only modestly slow activated dynamics far above the laboratory glass transition temperature.[35] Additional complications in PNCs include the presence of multiple interfacial and confinement effects,[10, 19, 30, 36] e.g., changes of entanglement density,[16, 17, 34, 37, 38] strong physical adsorption of polymer chains resulting in extremely long equilibration times[39] and pervasive nonequilibrium effects. These complications render it very difficult to definitively determine the existence or absence of glassy layers, spatial gradients of relaxation and elasticity, thickness of dynamically perturbed layers, and their temperature variations upon approaching $T_g$.

In this article, we employ the glycerol/silica nanocomposite (GSNC) as a model material in order to exclude the complications associated with chain confinement and entanglement effects, as well as the nonequilibrium chain adsorption phenomenon. The origin of matrix-particle attraction for miscibility is hydrogen bonding, which is representative of many practical PNCs. Based on a more rigorous data analysis approach that allows the model-free extraction of the full relaxation time distribution in the nanocomposite, our broadband dielectric spectroscopy (BDS) measurements show clear evidence for a dynamic interfacial layer. Key findings of the experimental analysis include a very broad distribution of (long) relaxation times, a mean slowing down of roughly one order of magnitude relative to the bulk, and a mean layer thickness that grows from ~1.8 nm at $T \sim 253$ K to ~3.5 nm near the bulk $T_g$ of glycerol. No indications of glassy layers are found from analysis of dielectric strength and temperature modulated DSC (TMDSC) data. In order to interpret our observations, and develop a real space understanding of spatial mobility gradients inaccessible to direct experimental measurement, we synergistically combine atomistic simulations with a lightly coarse grained statistical mechanical theory of activated relaxation that can bridge the high temperature simulation regime with the low temperatures relevant to experiment. Overall, we obtain good agreement between experiment, theory and simulation, and a detailed physical picture of the spatial mobility gradient in the interfacial layer is constructed. The validated theory is then employed to elucidate the role of material-specific local densification near the particle surface on slow dynamics and the formation of true glassy, high modulus layers.

## II. Details of experiments and simulations

### A. Sample preparations and characterizations

Pure glycerol, ethanol and methanol were purchased from Sigma-Aldrich and were used as received. The $SiO_2$ nanoparticles (NP) of radius of $R_{NP}$ =D/2= 12.5 nm were synthesized in ethanol at a concentration of 15 mg/ml by the modified Stöber method.[40, 41] The glycerol/$SiO_2$ nanocomposites (GSNCs) were prepared by the following procedure: First, 0.3 g of glycerol was dissolved in 10 ml ethanol. Then, different amounts of $SiO_2$/ethanol suspension were added into the glycerol/ethanol solutions in a drop wise manner while stirring. After one hour of mixing, the GSNCs were then dried in the hood at room temperature until most of the ethanol evaporated. After that, all the samples were further dried under vacuum conditions (~$10^{-5}$ bar) for another one week at 20 $^o$C. Good dispersion of GSNCs were achieved as evidenced by the sample transparency and TEM (Zeiss Libra 200 HT FE MC) after drying (see Fig. 1). Detailed sample characterizations are shown in Table 1. The loadings were measured by thermodynamic gravitation analysis (TGA) (Q50, TA Instruments). The volume fraction was calculated by assuming the density of glycerol is $\rho_{gly}$ = 1.26 g/cm$^3$ and the density of silica nanoparticles is $\rho_{sio_2}$ = 2.65 g/cm$^3$. The $T_g$ values of GSNCs were measured by temperature modulated differential scanning calorimetry (TMDSC) (Q1000, TA Instruments) from 0 °C to -100 °C with a cooling rate of 2 °C/min and modulation rate of $\pm 0.5$ °C/min. The average surface to surface interparticle spacing (IPS) $d_{IPS}$ was calculated as $d_{IPS} = R_{NP}\left(\left(\frac{3\varphi_f}{4\pi}\right)^{-1/3} - 2\right)$, where $\varphi_f$ is the volume fraction of nanoparticles. Broadband dielectric spectroscopy measurements for these nanocomposites in the frequency range between 10$^{-2}$ Hz to 10$^7$ Hz and temperature range between 273 K to 153 K were carried out using a Novocontrol Concept-80 system with an Alpha-A impedance analyzer, a Quatro Cryosystem temperature controller and a ZGS sample cell.

Table 1. Characteristics of glycerol/silica composites

| Sample | SiO$_2$ (wt%) | SiO$_2$ (vol%) | $T_g$ (TMDSC) (K) | $T_g$ (BDS) (K) | $d_{IPS}$ (nm) |
|---|---|---|---|---|---|
| GSNC 0 | 0 | 0 | 194 | 190 | -- |
| GSNC 8.8 | 16.9 | 8.8 | 195 | 190 | 20.3 |
| GSNC 15.6 | 27.9 | 15.6 | 196 | 190 | 12.4 |
| GSNC 23.6 | 39.4 | 23.6 | 196 | 190 | 7.61 |

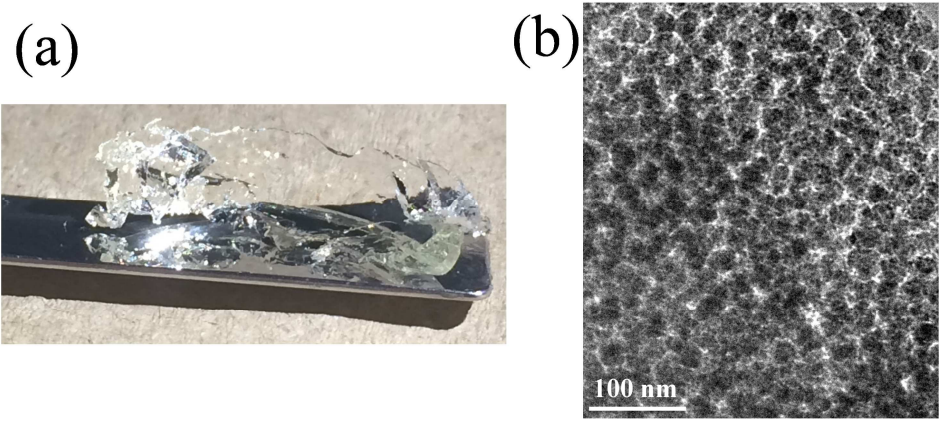

**Figure 1**: (a) Photograph of the glycerol/silica nanocomposite with loading of 23.6 vol%. (b) TEM image of the same composite to show the dispersion of nanoparticles. Some part of the image involves multilayers due to the difficulty of cryomicrotoming. The nanoparticles are individually seen and the sample is transparent, implying a good dispersion of nanoparticles.

### B. Dielectric responses of heterogeneous materials

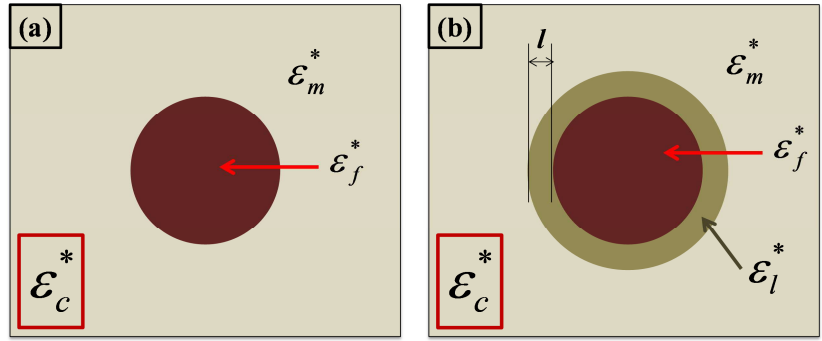

**Figure 2**: Geometry of the two-phase model (TPM) (a) and the interfacial layer model (ILM) (b), the overall dielectric response at a given frequency ω of the heterogeneous materials is $\varepsilon_c^*(\omega)$.

Dielectric spectra of nanocomposites are often analyzed as a superposition of spectra of different components. This approach is not accurate, because the dielectric responses of heterogeneous materials always contain contributions from the interference between different components[42, 43]. For example, the simplest two-phase model (TPM) with spherical particles (the sample geometry shown in Figure 2a) predicts the overall dielectric response $\varepsilon_c^*(\omega)$ at frequency $\omega$ calculated from Maxwell's equations[42] by assuming the effective medium boundary conditions:

$$\begin{aligned}
\varepsilon_c^*(\omega) &= \varepsilon_m^* \frac{(1+2\varphi_f)\varepsilon_f^*(\omega)+(2-2\varphi_f)\varepsilon_m^*(\omega)}{(1-\varphi_f)\varepsilon_f^*(\omega)+(2+\varphi_f)\varepsilon_m^*(\omega)} \\
&= (1-\varphi_f)\varepsilon_m^*(\omega) + \varphi_f \varepsilon_f^*(\omega) - (1-\varphi_f)\varphi_f * \frac{(\varepsilon_m^*(\omega)-\varepsilon_f^*(\omega))^2}{(1-\varphi_f)\varepsilon_f^*(\omega)+(2+\varphi_f)\varepsilon_m^*(\omega)} \\
&= (1-\varphi_f)\varepsilon_m^*(\omega) + \varphi_f \varepsilon_f^*(\omega) - (1-\varphi_f)\varphi_f * \varepsilon_{cross}^*(\omega)
\end{aligned} \quad (1)$$

where $\varphi_f$ is the volume fraction of nanoparticles; $\varepsilon_c^*(\omega)$, $\varepsilon_m^*(\omega)$ and $\varepsilon_f^*(\omega)$ are complex dielectric functions of the composite, the matrix and the nanoparticle, respectively. Thus the spectrum has not only contributions from two components with their volume fractions, but also a cross term that decreases the overall permittivity.

If an interfacial layer with dielectric function $\varepsilon_l^*(\omega)$ was added into the geometry (Figure 2b), the interfacial layer model (ILM) predicts even more complex dielectric function[43]:

$$\varepsilon_c^*(\omega) = \frac{\varepsilon_f^*(\omega)\varphi_f + \varepsilon_l^*(\omega)\varphi_l R^* + \varepsilon_m^*(\omega)\varphi_m S^*}{\varphi_f + \varphi_l R^* + \varphi_m S^*} \quad (2)$$

with $R^* = \dfrac{2\varepsilon_l^*(\omega)+\varepsilon_f^*(\omega)}{3\varepsilon_l^*(\omega)}$

$S^* = \dfrac{(\varepsilon_l^*(\omega)+2\varepsilon_m^*(\omega))(\varepsilon_f^*(\omega)+2\varepsilon_l^*(\omega))+2d(\varepsilon_l^*(\omega)-\varepsilon_m^*(\omega))(\varepsilon_f^*(\omega)-\varepsilon_l^*(\omega))}{9\varepsilon_l^*(\omega)\varepsilon_m^*(\omega)}$

$d = \dfrac{\varphi_f}{\varphi_f + \varphi_l}$

$\varphi_f + \varphi_l + \varphi_m = 1$

where $\varphi_l$ and $\varphi_m$ are the volume fraction of the interfacial layer and the matrix, respectively.

The above heterogeneous models provide the first order approximation for the overall dielectric response of our nanocomposites with spherical SiO$_2$ nanoparticles. In the analysis described below, $\varepsilon_f^* = 3.9$ holds in our experimental temperature and frequency range and the matrix $\varepsilon_m^*(\omega)$ is directly obtained from experiment on pure glycerol with a modified dc-conductivity to match the slightly higher dc-conductivity of our nanocomposites due to tiny amount of impurities from the NPs.

## C. Computer simulations

Atomistic molecular dynamics simulations of pure glycerol liquid and glycerol in contact with an amorphous silica substrate [44] were carried out using the class I force field, GAFF [45] and the MD code, LAMMPS [46] with GPU acceleration[47]. Details of the force-field and the comparison of the simulation results for pure glycerol to published density data,[48] the static structure factor[49] and the intermediate scattering function [50] of published neutron scattering experiments are provided in Appendix A.

In our simulations, 5000 glycerol molecules are placed in contact with the substrate. The center-of-mass of the substrate was tethered at the origin by a harmonic potential with interaction strength of 2000 kcal/mol. The simulation box is periodic in $x,y$ and $z$ directions and the silica substrate is replicated in the $xy$ plane (see Fig. A6). Initially, the system was equilibrated in the isothermal-isobaric (NPT) ensemble using a Nosé-Hoover thermostat and barostat, where the barostat was only applied in the direction perpendicular to the substrate. The NPT equilibration proceeded up to 5 ns followed by a NVT ensemble run for 100 ns. We tracked the parallel mean-squared-displacement (MSD$_\parallel$) of the center-of-mass (COM) of each glycerol molecule and averaged its MSD as a function of distance from the silica substrate, $z$.

A bin size of 0.25 nm was used and the counting ensured that both the initial and final position of the COM of the molecule belonged to the same bin. At long elapsed time, $MSD_{||}(z) \approx 4D_{||}(z)\Delta t$ such that the relaxation time at $z$ is computed as $\tau(z) \approx 1/D_{||}(z)$ and $\tau(z)/\tau_\alpha = D_{||,bulk}/D_{||}(z)$. To match the experimental nanoparticle loading of 23.6%, $<\tau g(\tau)>$ is computed based on data truncated at 3.8 nm from the surface of the substrate to match the experimentally determined IPS.

## III. Elastic Collective Nonlinear Langevin Equation (ECNLE) theory

Relevant technical aspects of prior ECNLE theory work[29, 51-53] in the bulk and free standing films are summarized in Appendix B. Here, we only recall the physical ideas. We then discuss the new challenges associated with solid surfaces and nanocomposites, and outline the zeroth order approach used in this article.

### A. Bulk Liquids and Free Standing Films

The foundational quantity to describe single molecule activated relaxation in the bulk liquid (volume fraction $\phi$) is a particle-displacement-dependent ($r$) dynamic free energy, $F_{dyn}(r) = F_{ideal}(r) + F_{caging}(r)$, which determines the effective force exerted on a moving tagged (spherical) particle due to its surroundings.[53] The localizing "caging" contribution, $F_{caging}(r;\phi,S(k))$, captures the effect of interparticle interactions and local structure on the nearest neighbor length scale ($r < r_{cage} \approx 3d/2$), and is quantified by the pair correlation function, $g(r)$, or Fourier space static structure factor $S(k)$. To execute a large amplitude local jump over the cage-scale barrier ($F_B$) as predicted from the dynamic free energy requires a small amount of extra space be created which is realized via a spontaneous collective elastic fluctuation of the liquid molecules outside the cage. The corresponding harmonic strain field decays as an inverse square power law of distance, and its amplitude is determined by the predicted jump distance. The resultant elastic barrier ($F_{elastic}$) is then determined by the harmonic stiffness of the transient localized state and the effective jump distance. The total barrier for the activated event is thus $F_{total} = F_B + F_{elastic}$.

To treat molecular liquids, a priori mapping to an effective hard sphere fluid is employed based on the requiring that the chemistry and temperature-dependent volume fraction *exactly* reproduces an equilibrium property of the real liquid, the dimensionless compressibility.[52] The latter quantifies the amplitude of nm-scale density fluctuations which is the key structural order parameter of ECNLE theory.

Using the above elements, Kramers theory is utilized to compute the mean barrier hopping time which plays the role of the alpha or structural relaxation time.[53] No adjustable parameter applications of the theory to molecular liquids, including glycerol, reveals good agreement with experiments over 14 decades in time.[52, 54]

For a free standing film, the bulk theory is modified because of two distinct, but coupled, physical effects.[29] First, the local barrier is lower near the surface due to a loss of nearest neighbors, the fraction of which can be analytically computed as a function of distance from the cage center, $\alpha(z)$. The dynamic free energy is thus:

$$F_{dyn}(r;z) = F_{ideal}(r) + \alpha(z) F_{caging}(r;\phi) \quad (3)$$

Second, the elastic cost for the re-arrangement is reduced since no strain field is present beyond the film interface. The presence of $\alpha(z)$ implies all physical quantities become position-dependent in the film, and as a consequence the elastic cutoff effect is coupled to the softened (for free interfaces) liquid near the surface and thus reduced caging effects are felt well into the film. The theory has been applied to predict mobility gradients, fast surface diffusion, $T_g$ shifts, and other properties of free standing films which have been favorably compared to experiment[29].

**B. Capped Films and Nanocomposites**

Solid surfaces introduce many new complications that are difficult to theoretically treat. Here, we present a first attempt which has the virtue of allowing quantitative predictions to be made that can be confronted with simulation and experiment.

The three key theoretical simplifications are as follows. (1) Liquids layer near a solid surface, and their local intermolecular structure also can change. Given the elementary length scale in bulk ECNLE theory is the cage radius, such packing changes should be incorporated on this scale. A typical manifestation of an attractive substrate is an increase of local density in *just* the first layer (thickness, *d*) of molecules[51] by a nonuniversal factor of $\lambda > 1$ compared to the bulk which depends on many physical and chemical factors. It modifies both the "effective" number of nearest neighbors of matrix molecules and the equilibrium pair correlations (or *S(k)*) as a function of distance from the surface. (2) A tagged particle near a solid wall is missing matrix nearest neighbors, but experiences forces due to surface atoms which will tend to (approximately) compensate for this loss. We assume the net effect of these competing force effects in the determination of the dynamic free energy is solely to densify the first layer of molecules. (3) The substrate atoms in real materials have variable degrees of vibrational motion or elastic stiffness. We assume that this motion is not perturbed during a matrix relaxation event. Hence, there is no elastic energy penalty inside the solid substrate, and the cutoff of the strain field idea formulated for free standing films is adopted. This simplest (and crude) assumption does not require any extra parameter to implement. Figure B1 shows that the increase in the *local barrier* due to near surface densification is predicted to be the primary reason that relaxation slows down near the solid surface.

Given the above simplifications, the dynamic free energy as a function of distance of a particle from the solid interface, $z$, depends on a position-dependent volume fraction, $\phi(z)$, as:

$$F_{dyn}(r;z) = F_{ideal}(r) + \alpha(z)F_{caging}(r;S(k;\phi(z))) \qquad (4)$$

Here, $\alpha(z) = \phi(z)/\phi_{bulk}$ quantifies the local densification by a factor of $\lambda > 1$, and can be analytically computed (see Appendix B). The position-dependent changes of the effective hard sphere fluid structure factor, $S(k;\phi(z))$, are computed using Percus-Yevick theory. As elaborated on in Appendix B, the cage scale barrier, $F_B$, is increased near the adsorbing surface out to a distance $r_{cage}+d$ from the interface. The amplitude of the elastic strain field and liquid stiffness associated with the nonlocal component of the alpha relaxation event, and hence collective elastic barrier, are again functions of the position in the confining geometry, and quantified based on idea (3).

The relaxation time spatial gradient then follows from Kramers theory.[53] From this, a "slow layer" thickness is identified as the spatial region where the relaxation time is significantly above its bulk liquid value. We find that this length scale, $l_{layer}$, depends weakly on chemistry (e.g., $\lambda$), but is essentially independent of temperature. One can then compute an average slow interfacial relaxation time as

$$\tau_{inter} = \frac{1}{l_{layer}} \int_0^{l_{layer}} dz\, \tau_\alpha(z) \qquad (5)$$

To compute the film-averaged distribution of relaxation times, $g(\tau)$, from knowledge of the mobility gradient, the relaxation process in each film layer is treated as a Poisson process controlled entirely by the mean relaxation time $\tau_\alpha(z)$. The probability density of relaxation times is thus $P(\tau, \tau_\alpha(z)) = \frac{\tau}{\tau_\alpha(z)^2} e^{-\tau/\tau_\alpha(z)}$. The experimental dielectric data was normalized using an integral over $ln(\tau)$, and relevant theoretical quantity is thus $\tau P(\tau, \tau_\alpha(z))$. For a film with thickness $h$ one then has

$$\langle \tau g(\tau) \rangle = 1/h \int_0^h dz\, \tau P(\tau, \tau_\alpha(z)) \qquad (6)$$

The local shear modulus (in the absence of relaxation) at a location $z$ in the film, and the layer-averaged frequency-dependent modulus, are computed using the methods employed previously[51] (see Appendix B).

In our experimental system, the nanoparticle and $d_{IPS}$ length are large compared to glycerol molecule ($d = 0.5\sim0.6$ nm) and the computed mobility gradient spatial range. Moreover, $D>>d$ implies that locally the particle appears as a flat surface to the matrix molecules (ignoring atomic scale surface corrugation). Both features simplify theoretical analysis. This motivates the adoption of a well-known[10, 36] analogy between nanocomposites and capped films by treating the effects of randomly dispersed particles on the matrix as identical to that of a flat surface. Thus, for the purpose of describing glycerol dynamics, the nanocomposite is mapped to a supported, semi-infinite film. Average nanocomposite quantities are calculated over a region extending from the surface into the matrix a thickness, $h = d_{IPS}/2$. The exact choice of $h$ has no impact on the calculations since $h$ is larger than any relevant length scale.

Obviously the range of validity of the 3 key simplifying ideas formulated above requires much future study. For our glycerol-silica system, two specific aspects should be mentioned. First, our treatment does not explicitly account for glycerol-surface hydrogen bonding forces, but some of their consequences are taken into account. In the successful treatment of bulk supercooled glycerol[50] with ECNLE theory, hydrogen bonding enters only via its significant modification of the key order parameter of the theory, the amplitude of nm-scale liquid thermal density fluctuations.[52] This is in the spirit of simplification (1) which assumes the primary *dynamical* effect of hydrogen bonding is modification of local structure. Concerning simplification (3), recall that ECNLE theory couples the caging and the longer range elastic fluctuation aspects. Our assumption that the strain field is cut off at the surface might be plausible given the high modulus of silica renders it hard to deform via small displacements of glycerol molecules. Also, since the two barriers for relaxation are coupled, the local densification parameter $\lambda$ affects both of them, albeit in different ways (see Fig.B1), and their relative importance will depend on system chemistry.

## IV. Results and Discussions

## A. Dielectric spectra of nanocomposites with different loadings

Figure 3 shows the experimental dielectric loss spectra of nanocomposites with different loadings in comparison with GSNC0. There are several notable features: (1) a significant increase of broadening in alpha relaxation peak upon increasing loading; (2) a reduction in alpha peak intensity relative to pure glycerol; and (3) a higher conductivity which may arise from a very small amount of impurities associated with the added nanoparticles. All these effects are similar to results reported for PNCs.[17-21] The inset of Fig.3 shows the temperature dependence of the alpha relaxation time estimated from the peak position of the loss spectra $\omega_{peak}$ ($\tau_\alpha = 1/(\omega_{peak})$). Good agreement between our measurements and literature data[55] indicate our samples are free of solvent from sample preparations.

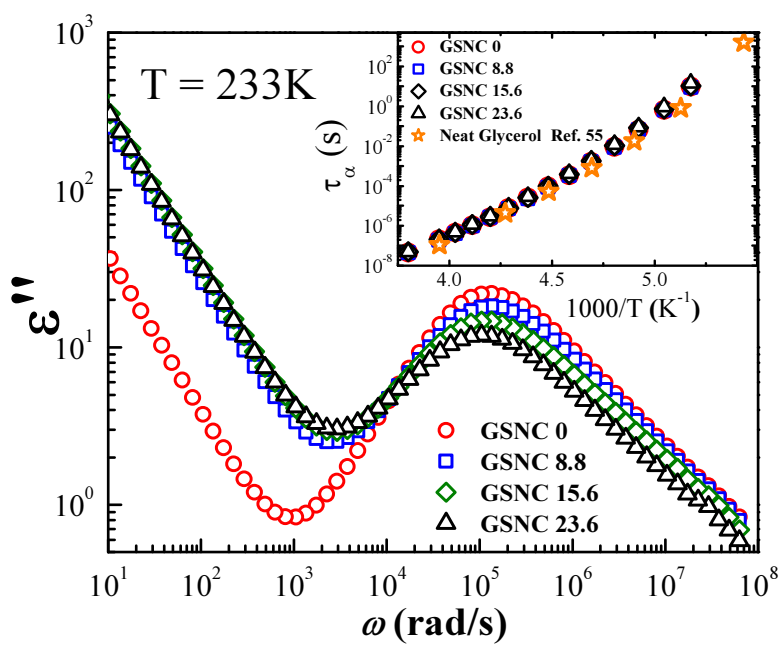

**Figure 3**: Dielectric loss spectra of composites with different loadings in comparison with neat glycerol at T = 233 K. The inset shows the alpha relaxation time estimated from the peak position of the loss spectra of our experiments and literature data of neat glycerol.[55]

## B. Signature of the interfacial layer dynamics with slowing down

**segmental relaxation**

Several methods have been proposed to interpret the dielectric broadening either by assuming an additional interfacial process contributes to the spectral broadening,[18, 20] or by hypothesizing that only free polymer is dielectrically active.[21] Such methods treat the spectra in an additive way, neglect the cross terms characteristic of heterogeneous systems (see section II.B), and are not accurate for quantitative analysis of the BDS spectra of a nanocomposite.[56] Here, we analyze our spectra in terms of the heterogeneous model (Section II.B) which provides a much more accurate analysis. As shown in Fig. 4, for a NP volume fraction ($\varphi_f$) of 23.6%, the TPM predicts a significant dielectric broadening (black line) even if the matrix dielectric relaxation is identical to pure glycerol. Thus, the observed spectral broadening can be ascribed partially to the heterogeneous nature of the nanocomposites.

On the other hand, the experimental spectra are clearly broader than predicted by the TPM model, which may indicate matrix dynamics do change upon the addition of nanoparticles.[19, 30, 36] Since $d_{IPS}>>d$ even at the highest loading studied, spatial confinement effects should be negligible. Hence, we attribute the difference between the TPM prediction and experiment to the existence of a dynamically perturbed glycerol layer near the NP surface. On the other hand, the interfacial layer model that includes a perturbed layer describes well the BDS spectra (both $\varepsilon'$ and $\varepsilon''$) over the entire frequency range.

We justify the above analysis by employing a model-free approach for calculating the relaxation time distribution directly from the dielectric function,[42] which can potentially separate two overlapping processes.[57] To rule out any NP contribution to the dielectric function, we back calculate, using the TPM, the effective matrix dielectric function $\varepsilon_{eff}^*(\omega)$ from the measured composite dielectric signal $\varepsilon_c^*(\omega)$. In this way, $\varepsilon_m^*(\omega)$ is replaced by $\varepsilon_{eff}^*(\omega)$ which contains only the matrix dynamics information. Figures 5a-5c show the real part $\varepsilon'_{eff}(\omega)$, derivative of the real part $\varepsilon'_{der}(\omega) = -\pi/2 * \partial\varepsilon'_{eff}/\partial\ln\omega$, and imaginary part $\varepsilon''_{eff}(\omega)$ of the effective dielectric spectrum $\varepsilon_{eff}^*(\omega)$ of the GSNC 8.8 and GSNC 23.6 samples. The neat glycerol spectra are presented for comparison. The $\alpha$-relaxation peak and the Maxwell-Wagner-Sillars (MWS) polarization peak can be clearly identified in the derivative spectra (Fig.5b). The MWS peak, according to the TPM, has the characteristic relaxation time

$$\tau_{MWS} = \varepsilon_0 \frac{(2+\varphi_f)\varepsilon_m + (1-\varphi_f)\varepsilon_f}{(2+\varphi_f)\sigma_m + (1-\varphi_f)\sigma_f}.$$

When $\sigma_m \gg \sigma_f$, $\varepsilon_m \sim 61$ and $\varepsilon_f \sim 3.9$, it appears at a frequency roughly of $f_{MWS} = \frac{\sigma_m}{2\pi\varepsilon_0\varepsilon_m}$, where the conductivity contribution to $\varepsilon''(\omega)$ becomes comparable to $\varepsilon'(\omega)$.[42, 58]

Since the $\varepsilon_{eff}^*(\omega)$ data contain no NP contributions, one can analyze the dynamics of the matrix more accurately from it. In linear response, the dielectric relaxation process can be described by a superposition of Debye functions with different relaxation times[42, 57]: $\varepsilon_{eff}^*(\omega) = \varepsilon_\infty + \Delta\varepsilon \int \frac{g(\ln\tau)}{1+i\omega\tau} d\ln\tau - i\frac{\sigma_0}{\varepsilon_0\omega^s}$, where $g(\ln\tau)$ is the distribution of relaxation times normalized as $\int g(\ln\tau)d\ln\tau = 1$, $\sigma_0$ is the dc-conductivity, $\varepsilon_0$ is the vacuum dielectric constant, the exponent $s$ is fixed to be 1.0 in our analysis, and $i$ is the imaginary unit.[57] We applied the generalized regularization method[59] to calculate this distribution as shown in Fig.5d, where the horizontal axis is intentionally reversed since $\tau = \omega^{-1}$.

Figure 5d demonstrates that only the $\alpha$-relaxation process is present in the $\Delta\varepsilon^*g(ln\tau)$ spectrum of pure glycerol, while the MWS process and an additional relaxation process appear in the nanocomposite spectra. The latter has a mean relaxation time ~10-20 times slower than the bulk $\alpha$-relaxation and corresponds to the low-frequency broadening of the spectra. We interpret it as direct evidence for the existence of a dynamically slower interfacial layer. To the best of our knowledge, this is the first clear experimental demonstration of the interfacial layer dynamics of a nanocomposite.

The overall relaxation time distribution shown in Fig.5d still contains relaxation processes from ions in the matrix: e.g. the MWS process and the dc-conductivity process, both of which are not relevant to the matrix segmental relaxation. Thus, we can subtract the MWS and conductivity contributions from the overall relaxation time distribution. As shown in Fig.6a, the temperature-dependent interfacial dynamics can be determined from the relaxation time spectra of all samples over the wide temperature range of 203-253 K. This distribution is a measure of the heterogeneous mobility gradient near nanoparticles, albeit with no spatial resolution. The maxima in the distribution functions are taken as the characteristic relaxation times of the $\alpha$-process ($\tau_\alpha$), and the shallow peak at longer times as the interfacial layer process ($\tau_{inter}$). Both $\tau_\alpha$ and $\tau_{inter}$ are essentially independent of NP loading (Fig.5d), as expected by experimental design. However, $\tau_{inter}$ does exhibit a slightly stronger temperature dependence than $\tau_\alpha$ (Fig.6a and inset of Fig.6b), with the ratio $\tau_{inter}/\tau_\alpha$ modestly increasing from ~ 8 to ~ 16 upon cooling for all nanocomposites. Significantly, there is a very broad tail of the distribution that extends to ~ 50 $\tau_\alpha$ with an amplitude that grows with cooling. Vogel-Fulcher-Tammann (VFT) fits of the mean relaxation times in Fig.6b suggest the BDS $T_g$ of the bulk matrix and interfacial layer are 190 K and 195 K, respectively, consistent with earlier studies of a thin glycerol film on a SiO$_2$ surface and under nanopore confinement.[60-62]

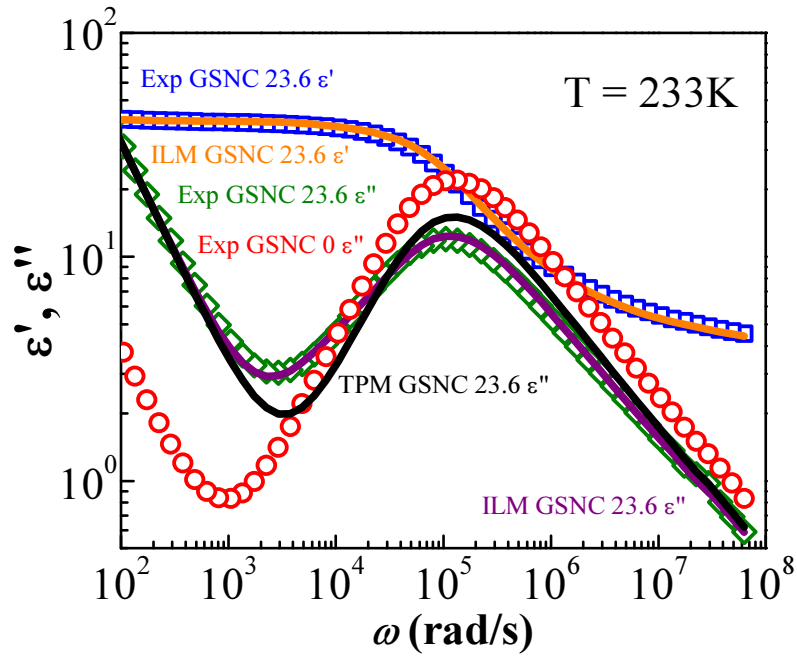

**Figure 4**: Raw dielectric spectra of GSNC 23.6 and pure glycerol at 233 K (symbols) as well as the spectra predicted by the TPM (black line) and the ILM (orange line for $\varepsilon'(\omega)$ and purple line for $\varepsilon''(\omega)$) at loading of 23.6%. The labels in the figure should be read as follows: **Exp GSNC 23.6 $\varepsilon$'** refers to the dielectric storage permitivity spectra $\varepsilon'(\omega)$ from **Exp**erimental measurements on the nanocomposite **GSNC 23.6**.

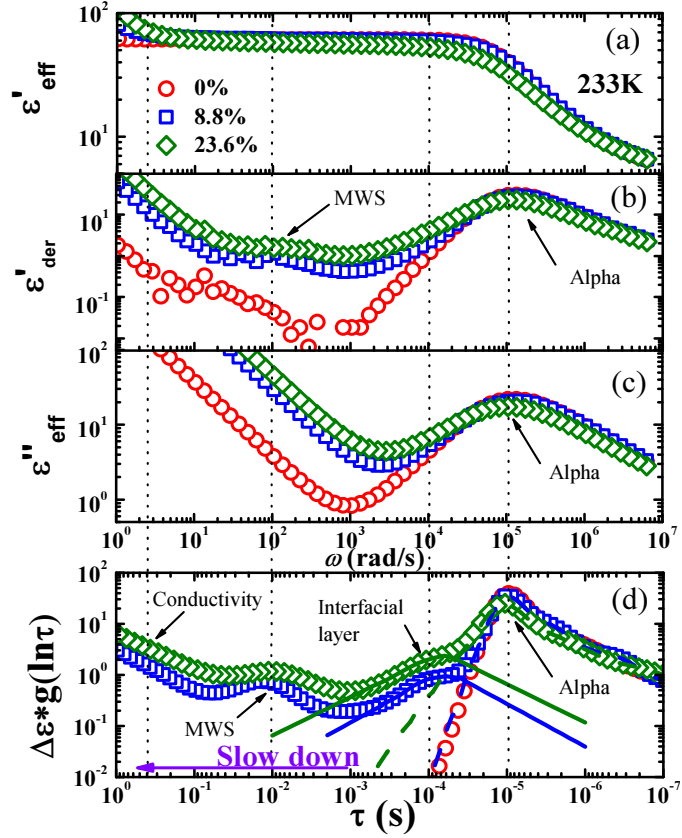

**Figure 5**: The effective dielectric function and relaxation time distribution from the same spectra of three different samples at 233 K: GSNC 0 (red circles), GSNC 8.8 (blue squares), GSNC 23.6 (green diamonds). (a) the real part $\varepsilon'_{eff}(\omega)$; (b) the derivative spectra $\varepsilon'_{der}(\omega) = -\pi/2*\partial\varepsilon'_{eff}/\partial ln\omega$; (c) the imaginary part $\varepsilon''_{eff}(\omega)$; (d) the relaxation time distribution $\Delta\varepsilon*g(ln\tau)$ of the corresponding samples. The solid curves indicate a fitting of the interfacial segmental process and the dashed curves show the fit of the bulk $\alpha$ peak of two nanocomposites.

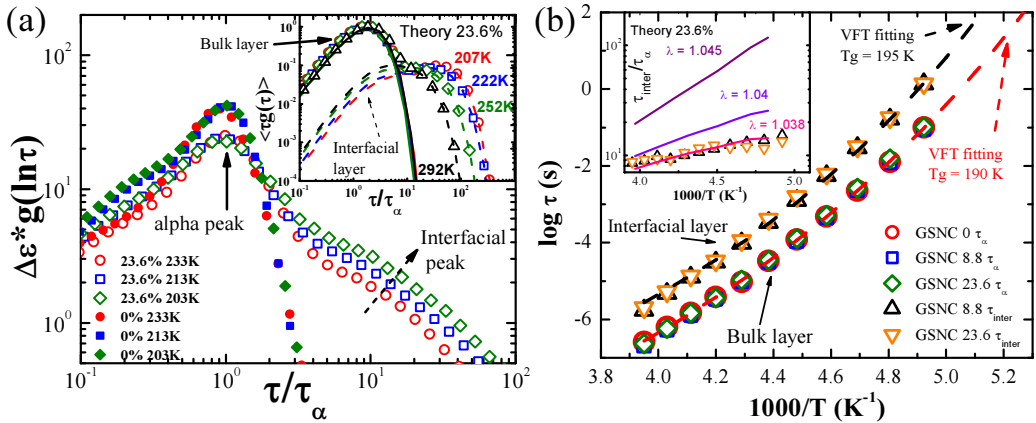

**Figure 6:** (a) The normalized distribution of experimental relaxation times at three different temperatures for pure glycerol (filled symbols) and nanocomposites with 23.6 vol% loading (empty symbols). The inset shows the theoretical prediction of the total relaxation time distribution (open symbols), the contribution from the bulk matrix (solid

curves) and the contribution from the interfacial layer (dashed curves) at four different temperatures 207 K (red), 222 K (blue), 252 K (green) and 292 K (black). As described in the text, the theory computes $\langle \tau g(\tau) \rangle = \left\langle \frac{\tau^2}{\tau_\alpha(z)^2} \exp(-\tau/\tau_\alpha(z)) \right\rangle$ . (b) Vogel-Fulcher-Tammann (VFT) fits (dash curves) of the bulk $\alpha$ relaxation time ($\tau_\alpha$) and interfacial layer $\alpha$ relaxation time ($\tau_{inter}$) suggest the BDS $T_g$ of the bulk and interfacial layer to be 190 K and 195 K, respectively. $\tau_\alpha$ and $\tau_{inter}$ are characteristic times from the relaxation time distribution analysis. The inset shows a comparison of the temperature dependence of the ratio $\tau_{inter}/\tau_\alpha$ from experiment and theoretical calculations performed at three different values of the local densification parameter $\lambda$. The empty symbols in the inset have the same meaning as in the main figure.

## C. Dynamics of the interfacial layer: comparison between experiments, theory and simulations

The nanocomposite dynamical measurements above exhibit many features of heterogeneous dynamics. Although some workers have suggested a mobility gradient near the nanoparticles from experiments and simulations,[15, 16, 32, 63] explicit experimental studies that clearly reveal this feature and provide a fundamental understanding of all features, especially in real space, are still missing. Moreover, no microscopic and quantitatively predictive theory currently exists to treat capped thin film and nanocomposites. Hence, our goals are 4-fold. (i) As described in section III and Appendix B, building on the ECNLE theory of bulk activated relaxation for molecular (including glycerol)[52-54] and polymeric liquids,[64] along with its successful generalization to free standing films,[29] we formulate a zeroth order theory for hopping relaxation in capped films and nanocomposites. (ii) To render the theory predictive for specific materials, we use results of our atomistic simulations (section IIC) at high temperatures where $\tau_\alpha$ < 100 ns. This provides local packing information required as an input to the dynamical theory, and serves as a benchmark to test the mobility gradient predictions in the lightly supercooled regime. (iii) The theory is then used to study the deeply supercooled regime inaccessible to simulation, compare against our experimental data, and to suggest the physical nature of the spatial mobility gradient that cannot be directly probed in experiments. (iv) The theory is applied to gain more generic insight concerning the connection between the material-specific densification at the particle-matrix interface and the degree of dynamical slowing, and elucidate the conditions required for forming glassy layers of specified shear elasticity.

Our theoretical predictions can be made without adjustable parameters if the local densification parameter $\lambda$ introduced in section IIIB is known. To obtain the latter in an a priori manner, we use results of atomistic simulations of the glycerol-silica system at relatively high temperatures T=293K. The microscopic equilibrium density profile shown in Fig.7a displays the familiar layering. Layer densities can be computed by integrating under the density profile, and are found to be equal to the bulk value (per the adopted theoretical model) for all layers except the first surface layer, where $\lambda \sim 1.04$ indicates denser packing.[65] The dynamic mean square displacements of glycerol molecules have also been computed, and the change of the mean $\tau_\alpha$ near the surface estimated. The same Poisson process model as described in section IIIB is employed to obtain the full distribution of relaxation times.

Figure 7b shows the gradient of reduced mobility at 293 K determined from simulation. An order of magnitude slowing down is found, and dynamics is suppressed over a few molecular diameters from the silica surface, but there is no "dead" layer ($\tau_\alpha$>100s). Figure 7b also shows ECNLE theory calculations of the mobility gradient using $\lambda$ = 1.038. The results are in good agreement with the simulation. This is especially significant since this value of $\lambda$ agrees (within the errors bars) with estimates from the equilibrium simulations. One can view this as a "calibration" step to encode chemically-specific information into the simplified structural model employed in the theory. In all subsequent comparisons of theory with our experimental data we fix $\lambda$ = 1.038. A good agreement between simulation and theory results for the distribution of relaxation times at 293 K (inset of Fig.7b) further supports the usefulness of the theory for local activated events that dominate at relatively high temperatures.

Theoretical calculations for how the mobility gradient evolves in the deeply supercooled regime are also shown in Fig.7b. Modest increase of the relaxation time near the particle surface relative to the bulk behavior is predicted with further cooling. However, the spatial range of the gradient of slow relaxation is essentially unchanged. Dissecting the theoretical calculations, and consistent with the 3 key simplifications invoked to formulate the theory in section IIIB, we find that the mobility gradient is largely determined by cage scale physics and near surface densification which enhances the local activation barrier. The small undershoot of the mobility gradient curve at low temperatures is a signature of cutting off the elastic field at the solid surface, a feature that is sensitive to the simplified theoretical approximation adopted.

The inset of Fig.7b shows that in the more deeply supercooled regime, theory and experiment are in overall good agreement for the full relaxation time distribution.

This includes both the breadth of the long relaxation time tail and its increasing amplitude with cooling. Quantitative deviations are evident deep into the high relaxation time tail. Use of the microscopic mobility gradient profile allows one to unambiguously separate the full relaxation time distribution into its perturbed layer and unperturbed components. Representative results are shown in the inset of Fig.6a. One sees not only shifting of the slow layer distribution to longer times with cooling, but significant broadening, modest amplitude growth, and other subtle shape changes that reflect details of the temperature-dependent real space gradient.

The inset of Fig.6b shows a comparison of theory and experiment for the ratio of interfacial layer mean alpha relaxation time with respect to bulk mean alpha relaxation time ($\tau_{inter}/\tau_\alpha$). Good agreement is obtained for the same value of $\lambda = 1.038$ deduced from comparison to our high temperature simulation. This provides additional support for the idea that the influence of a matrix-particle interface on relaxation can be captured reasonably well based on the local densification concept, at least for the glycerol composite. The inset also shows that the theoretical predictions for the magnitude and temperature dependence of the mean slowing down are remarkably sensitive to the chemically-specific degree of local densification. For example, increasing the latter from 3.8% to 4.5% results in a mean alpha time enhancement nearly 10 times larger near the bulk $T_g$.

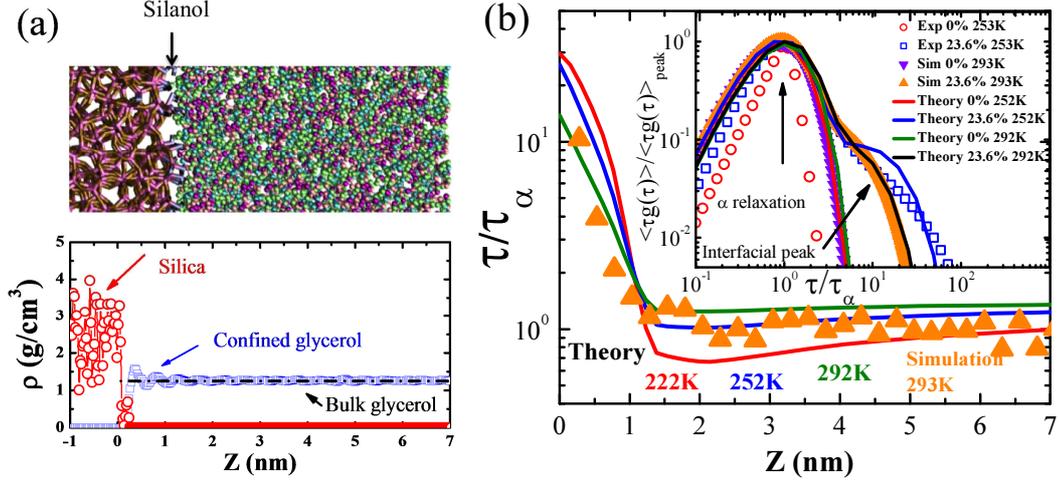

**Figure 7:** (a) Snapshot from atomistic molecular dynamics simulation of glycerol in contact with a SiO$_2$ substrate and its equilibrium density profile at $T$ =293 K; (b) The normalized (with respect to the bulk) $\alpha$ relaxation time gradient as a distance $Z$ from the surface as computed from simulation and theory (using $\lambda$ = 1.038). The different solid curves correspond to the theoretical mobility gradient at 222 K, 252 K and 292 K. The filled triangles are from simulations at $T$ = 293 K. The inset shows relaxation time spectra comparisons between theory, simulation and experiment, where theory and simulation plot $\langle \tau g(\tau) \rangle / \langle \tau g(\tau) \rangle_{peak}$ in comparison with $g(\ln\tau)/g(\ln\tau)_{peak}$ from experiments.

### D. Absence of glassy layers in glycerol/silica nanocomposites

From the above analysis of the segmental dynamics profile with spatial resolution, no indications of "glassy" dynamics ($\tau_\alpha$ > 100 s) were found. To determine the possible existence of a 'glassy' layer, we study in detail the dielectric relaxation strength, $\Delta\varepsilon$, and heat capacity, $C_p$, both of which are sensitive to the existence of the "glassy" layer.

For the dielectric relaxation strength, we emphasize that the crossterm (Eq.1) reduces the normalized dielectric strength, $\Delta\varepsilon_c/(1-\varphi_f)$, with increase in loading. This decrease can be clearly demonstrated in terms of the TPM:

when $\omega \to \infty$, $\varepsilon'_m(\infty) \sim \varepsilon'_f(\infty), \varepsilon'_c(\infty) \sim \varepsilon'_m(\infty)$;

when $\omega \to 0$, $\varepsilon'_m(0) \gg \varepsilon'_f(0)$, $\varepsilon'_c(0) \sim (1-\varphi_f)(1-\frac{\varphi_f}{2+\varphi_f})\varepsilon'_m(0) < (1-\varphi_f)\varepsilon'_m(0)$.

Therefore, $\Delta\varepsilon_c = \varepsilon'_c(0) - \varepsilon'_c(\infty) \sim (1-\varphi_f)\varepsilon'_m(0) - \varepsilon'_m(\infty) < (1-\varphi_f)\Delta\varepsilon_m$. In other words, the normalized dielectric strength, $\Delta\varepsilon_c/(1-\varphi_f)$, should not be a constant with loading, but rather should always be smaller than the dielectric strength of the neat glycerol $\Delta\varepsilon_m$ in our nanocomposites. The detailed analysis of experimental $\Delta\varepsilon_c$ (that includes both bulk-like and interfacial dynamics) and $\Delta\varepsilon_c/(1-\varphi_f)$ reveals good agreement with the predictions of the simple TPM for all loadings (Fig.8a), suggesting that all the glycerol dipoles relax within our frequency window. There is no sign of a measurable "glassy" or "dead" layer in the amplitude of the total dielectric signal.

Similar analysis can be done for the amplitude of the specific heat jump at $T_g$. The normalized specific heat capacity of the matrix, $C_p^{matrix} = \frac{C_p^{NC} - C_p^{NP} m_{NP}}{m_{matrix}}$, overlaps with that of neat glycerol, and the step feature in DSC is completed within 15K above $T_g$ for all the NP loadings, as shown in Fig. 8b. All these results conclusively demonstrate that there is no measurable glassy layer in our composites.

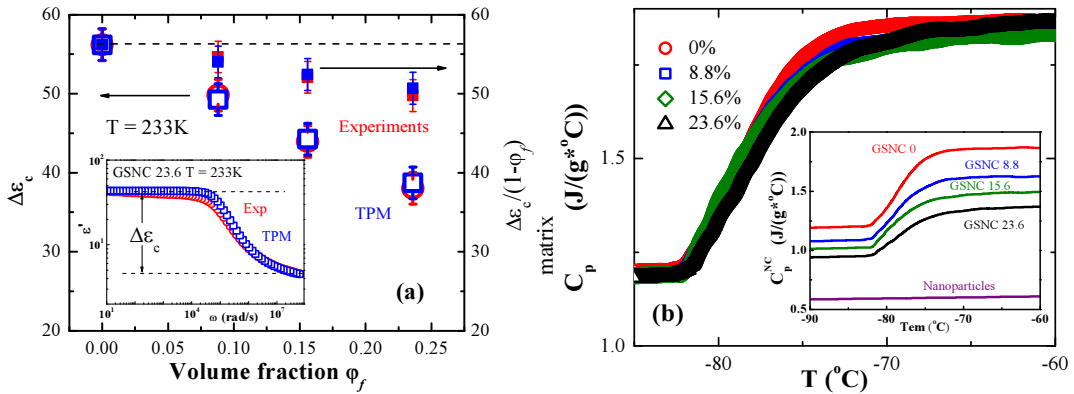

**Figure 8**: (a) Dielectric strength $\Delta\varepsilon_c$ (left axis) and normalized dielectric strength $\Delta\varepsilon_c/(1-\varphi_f)$ (right axis) at different loadings from experiments (red symbols) and TPM's

prediction (blue symbols), where $\Delta\varepsilon_c$ is the step increase of $\varepsilon'(\omega)$ and $\varphi_f$ is the volume fraction of the nanoparticles. The inset shows the estimates of $\Delta\varepsilon_c$ and the dielectric storage spectra $\varepsilon'(\omega)$ of GSNC 23.6 from experiments and the TPM. All the data are collected at T = 233K. (b) Normalized specific heat ($C_p^{matrix} = \dfrac{C_p^{NC} - C_p^{NP} m_{NP}}{m_{matrix}}$) of the glycerol matrix in Glycerol/SiO$_2$ nanocomposites and pure glycerol, where $C_p^{matrix}$, $C_p^{NC}$ and $C_p^{NP}$ are the specific heat of the matrix, nanocomposite and pure nanoparticle, respectively; $m_{matrix}$ and $m_{NP}$ are the mass fractions of matrix and nanoparticles, respectively. The inset shows the raw data of specific heat of pure glycerol, pure SiO$_2$ nanoparticles and nanocomposites with different loadings. No indications of a glassy layer are found in our nanocomposites according to TMDSC data.

### E. Interfacial layer thickness

Based on pre-averaging the spatial mobility gradient, the integrated dielectric spectral contributions in the ILM can be employed to deduce a mean interfacial layer volume fraction, $\varphi_l$. We find $\varphi_l$ increases from ~4% at 253 K to ~10% at 203 K for GSNC 8.8, and from ~12% to ~28% for GSNC 23.6. From this, the average interfacial layer thickness $l$ is estimated as $l = \left(\left(\dfrac{\varphi_l + \varphi_f}{\varphi_f}\right)^{1/3} - 1\right) R_{NP}$, yielding a thickness that grows from ~1.8 nm at $T$ = 253 K (~$T_g$ + 58 K) to ~3.5 nm at $T$ = 203 K (~$T_g$ + 8 K), as shown in the Fig.9. These values are (as expected) independent of nanoparticle volume fraction. Such a cooling-induced growth of the interfacial thickness is consistent with the direct observation of an intensity drop of the main loss peak (inset in Fig.9); e.g., $\Delta\varepsilon'' \sim 3.6$ at 253 K versus $\Delta\varepsilon'' \sim 7$ at 203 K, corresponding to a ~17% and ~28% decrease in $\Delta\varepsilon''$ relative to pure glycerol. Since $\Delta\varepsilon \sim \dfrac{N}{V} \sim \int \varepsilon''(\omega) d\ln\omega$, where $\Delta\varepsilon$ is proportional to the number density, $N/V$, of molecules belonging to a specific relaxation process and $\varepsilon''(\omega)$ is the loss spectra intensity,[42] the decrease of the $\varepsilon''(\omega)$ intensity can be taken as indirect evidence of an increase of the glycerol population in the slower interfacial layer which monotonically grows upon cooling. We emphasize that the estimated interfacial layer thickness is consistent with a glycerol thin film study[60] which found $l \sim 1.6–2.5$ nm at $T = 233$ K. A similar trend upon cooling was observed for deeply supercooled salol[66] under surface confinement and 2-methyltetrahydrofuran[67] under nanopore confinement. The change of the layer thickness is also consistent with the logarithmic growth of the number of dynamically correlated glycerol molecules in the same supercooled temperature region as estimated using nonlinear dielectric measurements.[68]

However, such an increase of the interfacial layer thickness is not captured by our zeroth order theory. As indicated in Fig.7b, both simulation and theory find a layer thickness of ~1.3-1.5 nm at high $T = 293$ K, consistent with experiment. Upon cooling, the theory does not predict that the growth of the layer of retarded relaxation based on the real space mobility gradient. In contrast, experiments deduce a growing layer thickness with cooling, which is partially based on the increase of the long time tail amplitude in the relaxation time distribution in Fig.6a. The implications of this apparent disagreement is unclear given the theory *does* correctly predict the relaxation time distribution amplitude grows with cooling (insets of Fig.6a and 7b). One possibility is our minimalist assumption (simplification (3) in section IIIB) that the elastic contribution to the collective barrier is fully cut off at the solid surface is not quantitatively accurate at low temperatures. If distortion of the solid substrate is required for the matrix relaxation, we expect the elastic barrier will grow relative to its bulk value with cooling resulting in an enhanced range of the spatial mobility gradient. The theory also ignores surface corrugation,[66] possible structural ordering,[69] and perhaps other aspects of cooperative motion,[32] factors which might become relevant at low temperature.[66]

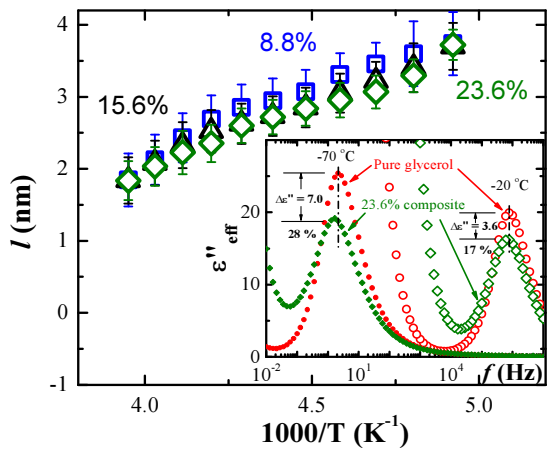

**Figure 9:** Average interfacial thickness as a function of temperature based on the ILM for all three samples: 8.8% loading (squares), 15.6% loading (triangles) and 23.6% loading (diamonds). The inset shows a direct comparison of $\varepsilon''_{eff}(\omega)$ of pure glycerol and nanocomposites with 23.6% loading at two temperatures.

### F. Predicting the interfacial layer properties

The present zeroth order theory can be utilized to suggest guidelines for controlling the degree of slowing down and temperature dependence of interfacial relaxation, including the conditions required to realize kinetic glassy layers. Here, we perform model calculations of the mobility gradient, interfacial layer mean relaxation time and dynamic shear modulus, as a function of the chemically-specific degree of local densification $\lambda$, and reduced temperature $T/T_{g,bulk}$, or equivalently the pure matrix alpha relaxation time. We employ glycerol as a representative surrogate of molecular matrix materials. The layer dynamic elastic shear modulus at fixed external frequency is computed using the relations given in Appendix B.

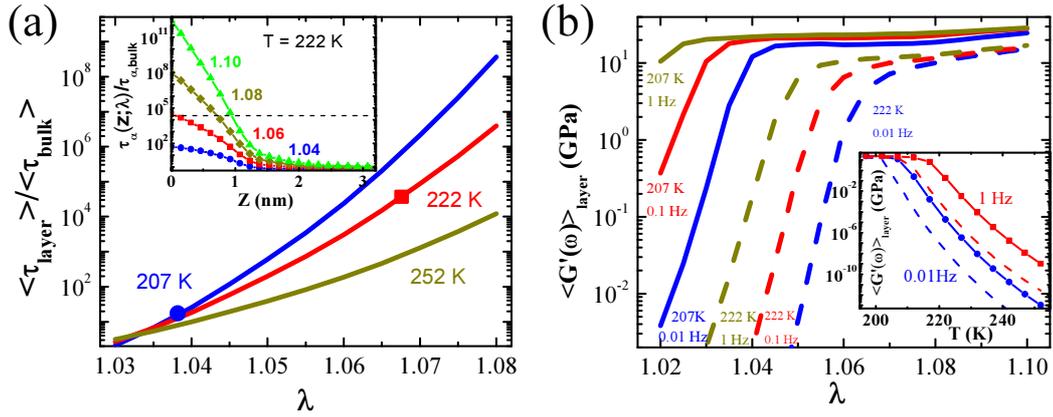

**Figure 10**: (a) Average relaxation time in the interfacial layer normalized by the bulk liquid relaxation time as a function of the densification parameter $\lambda$ for glycerol at 207 K (blue), 222 K (red), and 252 K (yellow). For reference, the theory predicts that the bulk $T_g$ is 202 K, and bulk relaxation time is approximately 5.8 s at 207 K, $2.7 \times 10^{-3}$ s at 222 K, and $5.2 \times 10^{-7}$ s at 252 K. The points marked on the lower temperature curves indicate the value $\lambda = \lambda_g$ where the interfacial layer mean relaxation time becomes 100 s representing one measure of the transition to a "glassy layer". The inset shows the relaxation time profile near the surface (in units of its bulk analog) at 222 K for $\lambda = 1.04$ (blue circles), 1.06 (red squares), 1.08 (yellow diamonds), and 1.1 (green triangles). The dashed line represents 100 s and serves to show that the existence of a glassy layer above a critical $\lambda_g$ and its growth in thickness with increasing $\lambda$. (b) Average dynamic shear modulus in the interfacial layer of glycerol as a function of the densification parameter $\lambda$ at frequencies $\omega = 0.01$ (blue), 0.1 (red), and 1 (yellow) Hz. The solid (dashed) curves are at 207 K (222 K). The theoretical dynamic moduli for bulk glycerol are: (i) at 0.01 Hz, 0.055 GPa at 207 K and 19 Pa at 222 K, (ii) at 0.1 Hz, 4.1 GPa at 207 K and 1.9 kPa at 222 K, (iii) at 1 Hz, 15.6 GPa at 207 K and 0.19 MPa at 222 K. The inset shows the average dynamic shear modulus in the interfacial layer with $\lambda = 1.04$ at $\omega = 0.01$ (blue circles), 1 (red squares) Hz. The dashed lines of corresponding color show the modulus in bulk glycerol at the same frequency.

The main frame of Fig.10a shows the theoretical mean layer relaxation time relative to its bulk analog as a function of the densification parameter at three temperatures. The latter are chosen to roughly correspond to the onset of the deeply supercooled regime ( $\tau_\alpha = 520 ns$ ), middle of the deeply supercooled regime ( $\tau_\alpha = 2.7 ms$ ), and close to the bulk $T_g$ ( $\tau_\alpha = 5.8 s$ ). One sees a strong, *supra*-exponential growth of the interfacial layer relaxation time with increasing surface densification parameter, which becomes more dramatic at lower temperatures.

At 207 K and 222 K, glassy layers emerge above the bulk $T_g$ beyond a critical value of surface densification. The inset shows examples of the full mobility gradients at a fixed temperature of 222K (bulk $\tau_\alpha = 2.7 ms$) for four values of $\lambda$. As local densification grows from 4% to 10%, the relaxation time near the surface sharply increases by 10 orders of magnitude. The thickness of the "glassy layer" (when it exists) grows with increasing surface densification, and is ~ 1 nm at $\lambda$ = 1.1. The overall interfacial layer thickness, defined as the distance from the surface required to recover the bulk relaxation time, also grows monotonically with $\lambda$ from ~ 1.5 nm to ~ 3 nm (~2.5-5 glycerol diameters).

Figure 10b shows representative interfacial layer dynamic shear modulus calculations as a function of $\lambda$ at two temperatures (207 K, 222 K) and three (low) frequencies of 0.01, 0.1 and 1 Hz. The corresponding shear moduli of bulk glycerol are stated in the caption. At 0.01 Hz, with increasing surface densification the interfacial layer shear modulus grows by over 2 decades at 207 K, and by nearly 9 orders of magnitude at 222 K. At both temperatures, and all three frequencies studied, a glass-like modulus of 5-10 GPa is attained at sufficiently high local surface densification. At fixed temperature and $\lambda$, layer shear moduli grow with increasing frequency, as expected. The inset shows an example at two frequencies of how increasing interfacial layer shear rigidity emerges upon cooling at fixed surface densification ($\lambda$ = 1.04). Massive increases are found in all cases, following a roughly identical temperature-dependent form. The calculations suggest that the average mechanical stiffness of the layer can be tuned from liquid-like to glass-like over a narrow range of temperature.

## V. Conclusions

We have combined broadband dielectric spectroscopy, atomistic simulation, and a novel activated relaxation theory to unravel the nature of slow dynamics in the interfacial layer of a model glycerol-silica nanocomposite. Due to its relative simplicity, the glycerol-silica system serves as a model system to study the complex interfacial effects in nanocomposites, and the new data analysis methods presented here can be applied to probe dynamics in more complex polymer nanocomposites. Our analysis reveals that the relaxation near the particle surface slows down relative to bulk behavior, and increasingly so upon cooling. We ascribe the observed slowing down of the interfacial dynamics mainly to the densification of the liquid near the attractive surface and its influence on the cage scale local barrier to activated hopping. The thickness of the interfacial layer increases upon approaching the bulk $T_g$. However, no indications of glassy layers are found from analysis of dielectric relaxation strength or from the step feature in the DSC measurements. In addition, we have formulated a zeroth order statistical mechanical theory of activated, spatially-resolved interfacial dynamics near a solid surface under lightly and deeply supercooled conditions. Once the near surface densification parameter, required as input to the theory, is deduced from equilibrium atomistic simulations at high temperature, the theory provides quantitative predictions for dynamics and mechanical properties of the interfacial region. Estimates of the densification parameters required for the existence of a true glassy layer were also made. Encouraging agreement between experiment, theory and simulation has been obtained. However, we emphasize that much future theoretical work is required to deeply test the range of applicability of our simplifications, and to improve the theory by taking into account the many complicating features present in real nanocomposite materials.

**Appendix A: Simulations Details**

We performed atomistic molecular dynamics (MD) simulations for pure glycerol liquid and glycerol in contact with an amorphous silica substrate. The simulations for pure glycerol were used to calibrate the force field through comparison with published experimental data on density at different temperatures and with neutron scattering results (e.g., structure factor and intermediate scattering function). The simulations for glycerol on a model silica surface are based on the calibrated force field.

To model the intra- and inter-molecular interactions of glycerol we used the Generalized Amber Force Field (GAFF).[45] The total potential energy of the system $U_{TOTAL}$, consists of the non-bonded interaction, bond, angle, and dihedral potentials:

$$U_{TOTAL} = \sum_{i<j}\left[\frac{A_{ij}}{R_{ij}^{12}} - \frac{B_{ij}}{R_{ij}^6} + \frac{\delta_i \delta_j}{\varepsilon R_{ij}}\right] + \sum_{BONDS} K_r(r-r_{eq})^2 + \sum_{ANGLES} K_\theta(\theta-\theta_{eq})^2 + \sum_{DIHEDRALS} K_{\phi_D}[1+\cos(n\phi_D)] \quad (A1)$$

where $r_{eq}$, and $\theta_{eq}$ are equilibrium structural parameters, $K_r$, $K_\theta$, and $K_{\phi D}$ are force constants, $n$ is multiplicity, and $A$, $B$ and $\delta$ are parameters that characterize the non-bonded potentials. In this force field, the 1-4 non-bonded interactions (e.g., non-bonded interactions for atoms that are connected 3 bonds apart such as the end atoms in a dihedral) are weighted at a fraction of 1/2 for the van der Waals interaction and 5/6 for Coulombic interactions. The van der Waals cross-term is the geometric mean of the components for the interaction potentials while the van der Waals radius is the arithmetic mean of the components. In evaluating the long-range electrostatic interactions we used the standard PPPM algorithm implemented in LAMMPS with an accuracy of $1 \times 10^{-4}$ and a near field cutoff set at 10 Å.

For the neat glycerol system, the MD simulation consisted of 1000 molecules, placed in a simulation box under periodic boundary conditions. The system was equilibrated at isothermal-isobaric (NPT) ensemble conditions using a Nosé-Hoover thermostat and barostat. An integration time step of 1 fs was used to solve the equation-of-motion. The set point for the thermostat was fixed at specific temperatures that correspond to experimental studies in the literature ($T$ = 273.15 K, $T$ = 298.15 K or $T$ = 313.15 K) and the barostat was set to a pressure of 1 atm. The simulation was first equilibrated for 5 ns—where the pressure, temperature and volume fluctuations of the simulation had stabilized—and then followed by a production run for 100 ns in the canonical (NVT) ensemble at the same temperature and average volume as the NPT run.

The partial charges, $\delta$, of each atom in a glycerol molecule are shown in Fig.A1. To obtain these values we optimized the density of glycerol as a function of temperature using molecular dynamics simulations so it reproduced the experimentally measured dependence over the temperature range of 273 K- 313 K (see Fig.A2). The initial values of $\delta_i$ were obtained from semi-empirical (AM1) with bond charge correction (BCC) calculation method that is included in the Antechamber[70] package. Since a glycerol molecule is neutral ($d_f \sum_i \delta_i = 0$), we optimized the partial charges by performing a parameter sweep of $d_f$ from 1.00 to 1.12. A value $d_f$ = 1.03 results in the best agreement in $T$ at 293.15 K and $P$ = 1 atm as shown in Fig.A2.

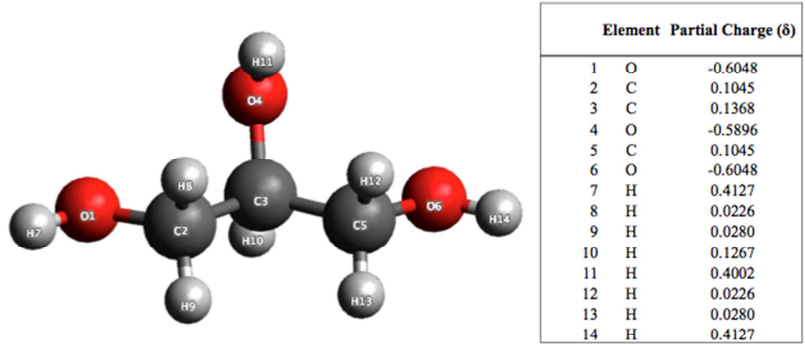

**Figure A1:** Partial charges, $\delta$, of a glycerol molecule used in our simulations.

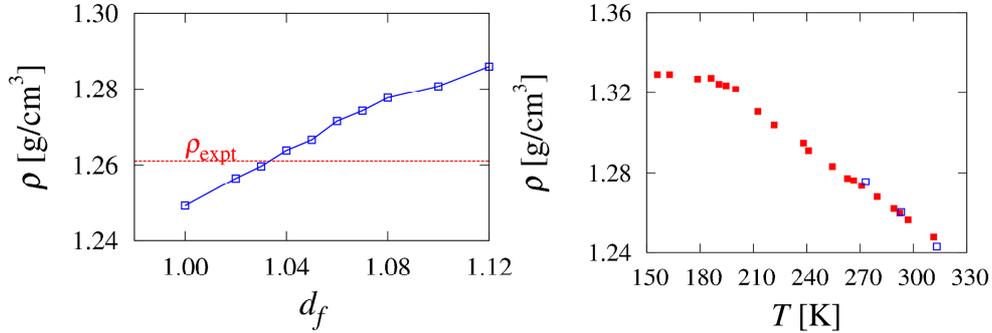

**Figure A2:** (Left) Density of the simulated glycerol for different values of $d_f$ at $T$ = 293.15 K and $P$ = 1 atm where $\rho_{expt}$ (red dotted-line) is the density obtained from experiments at 293 K and 1 atm.[46] (Right) Density of glycerol obtained from experiments[46] (filled-square) and from simulations with $d_f$ = 1.03 (open-square) for different temperatures and $P$ = 1 atm.

We compared the structure of the simulated glycerol solution with the neutron scattering experiments of Towey et al.[49] The collective static structure factor is defined as,

$$I(Q) = \frac{1}{V} \langle | \sum_{m=1}^{N} b_m \exp(-i\vec{Q} \cdot \vec{r}_m) | \rangle \qquad (A2)$$

where $b_m$ is the scattering length density, $\vec{Q}$ is the momentum wave vector, $\vec{r}$ is the position vector and $V$ is volume. Fig.A3 compares our simulated data with the experimental result. Relevant scattering peaks are captured and the simulation results are in overall excellent agreement with experiment.

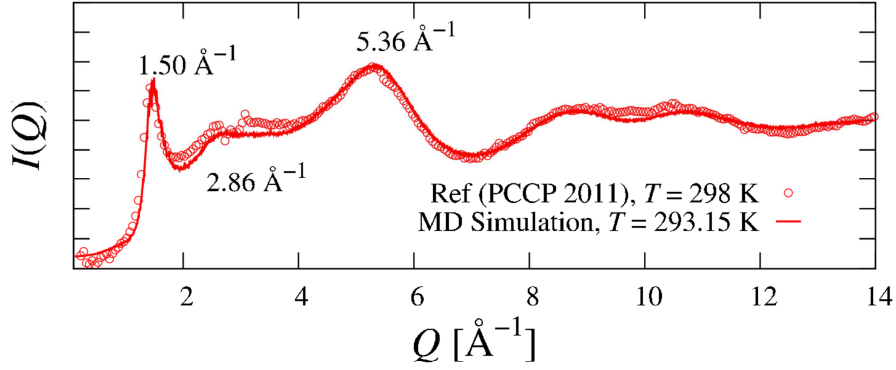

**Figure A3:** Static structure factor of glycerol.

We calculate the intermediate scattering function of the glycerol liquid from $\frac{S(Q,\Delta t)}{S(Q,0)} = \frac{\langle I(\vec{Q},\Delta t)\bullet -I(\vec{Q},t)\rangle}{\langle I(\vec{Q},t)\bullet -I(\vec{Q},t)\rangle}$ at $Q = 1.50$ Å$^{-1}$, which is the first peak of the static structure factor (see Fig.A4). This procedure is similar to what was done in the neutron scattering study of deuterated glycerol.[50] We then fit $\frac{S(Q,\Delta t)}{S(Q,0)}$ to a stretched exponential, $\frac{S(Q,\Delta t)}{S(Q,0)} = A\exp\left[-\left(\frac{t}{\tau_\rho}\right)^\beta\right]$ and obtained $\beta = 0.751$ and $\tau_\rho = 763$ ps. In comparison, the experimental values for 293 K are $\beta = 0.7$ and $\tau_\rho = 3465$ ps, which indicates the dynamics of our simulations are faster by a factor of 4.5.

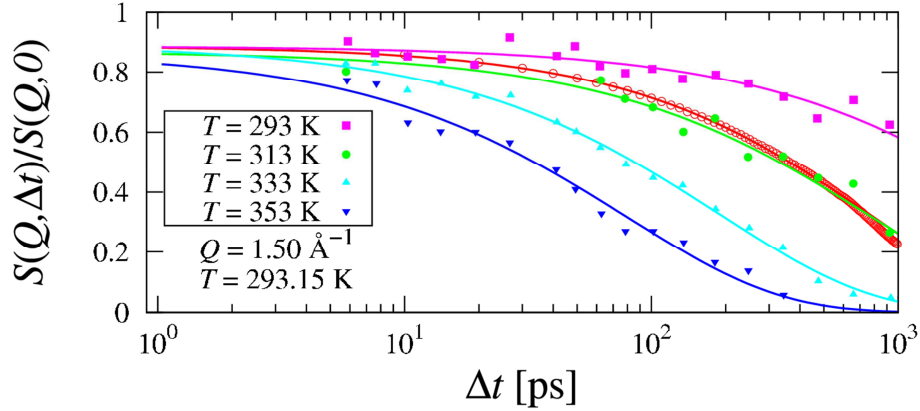

**Figure A4:** Intermediate scattering function of deuterated glycerol at $Q = 1.44$ Å$^{-1}$ at different temperatures (symbols inside the boxed legend) and from simulations at $Q = 1.50$ Å$^{-1}$. The lines are stretch exponential fits.

Further comparison between simulation and experiment is possible by comparing the self-diffusion coefficients. The bulk diffusion coefficient, $D_b$, of glycerol at 293 K is 0.000137 Å$^2$/ps.[71] In simulations the diffusion coefficient can be obtained from the mean-square displacement (MSD) of the center-of-mass (COM) using the equation:

$$MSD = a + 6D_b \Delta t \qquad (A3)$$

In Fig.A5 we have fit the MSD data of our simulation and found $D_b = 0.000590$ Å$^2$/ps. The ratio between the simulation and the experimental value is 4.3 and again indicates faster dynamics in the simulation. This ratio is in agreement with what was found using $\frac{S(Q,\Delta t)}{S(Q,0)}$.

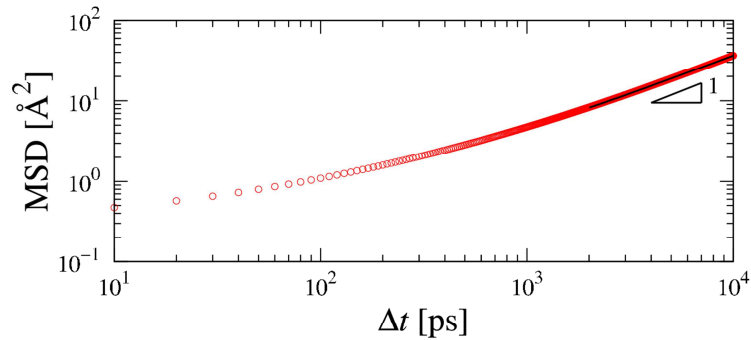

**Figure A5:** Mean-square-displacement of simulated glycerol.

In the simulation of glycerol in contact with a silica substrate, 5000 molecules of glycerol were placed in contact with the model silica substrate as shown in Figure A6. The center-of-mass of the substrate, ($x_{cm}$, $y_{cm}$, $z_{cm}$) was tethered at the origin, (0,0,0), by a harmonic spring with potential energy, $U_{spring} = \frac{K_s}{2}(x_{cm}^2 + y_{cm}^2 + z_{cm}^2)$, where $K_s$ = 2000 Kcal/mol. Initially, the system was equilibrated in the isothermal-isobaric (NPT) ensemble using a Nose-Hoover thermostat and barostat, where the barostat was only applied in the direction perpendicular to the substrate. The NPT equilibration proceeded up to 5 ns ($T$ = 293.15 K and $P$ = 1 atm) followed by a NVT ensemble run for 100 ns at $T$ = 293.15 K.

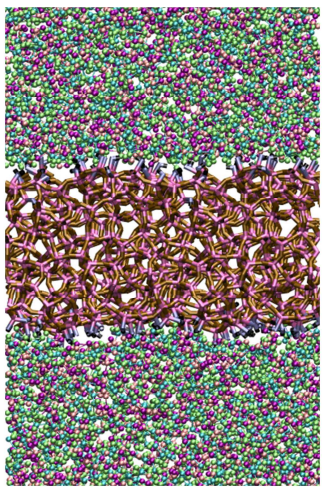

**Figure A6:** Snapshot of glycerol in contact with a model silica substrate (middle slab). The simulation box is periodic in $x,y$ and $z$ directions. The silica substrate bonds are bonded with its periodic image in both $x$ and $y$ directions. The image is truncated and ranged from $z$=-3.8 nm to $z$=3.8 nm for clarity.

In Fig.A7, the parallel component of the MSD, MSD$_\parallel$, which is used as raw data for Fig.7b of the main text, was calculated for different $z$ locations and compared with that of 2/3×MSD of pure glycerol (the 2/3 pre-factor adjusts the dimensional quantity). Furthermore, Fig.A7 shows MSD$_\parallel$ reaches Fickian diffusion on the timescale of nanoseconds even when z~2 Å (near the substrate and high temperature). However, this would not be the case for low-temperature simulations where the relaxation time is significantly greater than a typical atomistic MD simulation time.

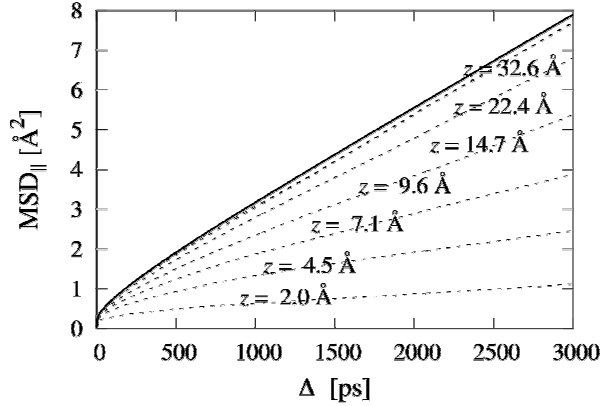

**Figure A7:** Parallel mean-square displacement, MSD$_\parallel$ at different $z$ locations (dotted-line) and for pure glycerol (solid-line).

**Appendix B: Details of ECNLE Theory**

The technical aspects of Elastically Cooperative Nonlinear Langevin Equation (ECNLE) theory for activated relaxation in the bulk and free-standing films composed of molecular liquids have been discussed at great length in the literature.[29, 51-53] In the bulk, the dynamic free energy experienced by a tagged particle in the liquid is[53]:

$$\beta F_{dyn}(r) = -3\ln(r) + \frac{6\phi}{\pi}\int \frac{d\vec{k}}{(2\pi)^3} \frac{C^2(k;\phi)S(k;\phi)}{1+S^{-1}(k;\phi)} \exp\left[-\frac{k^2 r^2}{6}(1+S^{-1}(k;\phi))\right]$$

$$= F_{ideal}(r) + F_{caging}(r;\phi) \qquad (B1)$$

where for spherical particles $S(k) = (1-\rho C(k))^{-1}$ is the collective static structure factor in Fourier (k) space, $\phi$ is the volume fraction, and all lengths are in units of the particle diameter $d$. The second caging term captures the effect of local interparticle interactions due to nearest neighbors on length scales $r < r_{cage} \approx 3d/2$. The dynamic free energy defines a local barrier height, $F_B$, and a jump length scale, $\Delta r \leq 0.4d$, both of which grow with cooling.[52, 53] To execute a large amplitude local jump requires a small cage expansion which is realized via a spontaneous collective elastic fluctuation of the liquid molecules outside the cage. The corresponding strain field, $\vec{u}(x)$, decays as an inverse square power law of distance, and its amplitude is determined by the jump distance. The resultant elastic activation barrier is[53]

$$F_{elastic} \approx \int_V d\vec{x}\rho\left(\frac{1}{2}u(\vec{x})^2\right)K_0 = 12\phi\Delta r_{eff}^2 r_{cage}^3 K_0 \qquad (B2)$$

Here, $\Delta r_{eff} \approx 3\Delta r^2/32 r_{cage}$, $K_0$ is the curvature of the minimum of the dynamic free energy that determines the transient vibrational amplitude, $\vec{x}$ is a vector with origin at the center of the cage region of the local relaxation event, and $V$ is the volume of the liquid outside the cage. The total barrier is the sum of local and collective elastic contributions, $F_{total} = F_B + F_{elastic}$. Molecular liquids are treated by mapping them to an effective hard sphere fluid based on the requiring that the mapping *exactly* reproduces the equilibrium dimensionless compressibility of the specific real liquid.[52] Kramers theory is then employed to compute the mean barrier hopping time which is taken as a faithful measure of the alpha or structural relaxation time.[53]

For a free standing film with two vapor interfaces [29], the local barrier is lowered close to the surface due to a loss of nearest neighbors, the fraction of which can be analytically computed as a function of distance of the center of the cage from the surface, $\alpha(z)$. The elastic cost for the re-arrangement is also reduced since the strain field is cut off at the film interface. The microscopic density profile is taken to be a step function with the bulk density up to the surface, and zero outside the film, and the vapor interface does not disturb liquid packing. The local barrier is a function of depth in the film, and its bulk value is recovered beginning at a distance $r_{cage}$ from the surface. The quantities $\vec{u(x)}$ and $K_0$ (harmonic spring constant describing the transient localized state) become functions of the position in the confining geometry and the analog of Eq.(B2) is

$$F_{elastic}(z) = \int_V d\vec{x} \left( \frac{1}{2} \rho u^2(\vec{x};z) K_0(\vec{x}) \right) \tag{B3}$$

where $z$ is the distance from the surface to the tagged particle. The volume integral extends only over the finite region defined by film geometry[29].

Our zeroth order generalization of the free standing film theory treats solid surfaces based on the 3 simplifications stated in the main text. Near a solid wall the local structure changes (point (1)), and the dynamic free energy is given by Eq.(4) of the main text. Using this dynamic free energy, where the fractional loss of neighbors as a function of distance from the surface follows from an elementary calculation as

$$\alpha(z) = \begin{cases} \dfrac{z^3 + 3(\lambda-1)z^2 - \left(3(\lambda-1)+3r_{cage}^2\right)z - \left(3r_{cage}^3 + 3r_{cage}^2(\lambda-1) - (\lambda-1)\right)}{(z+r_{cage})^2(z-2r_{cage})} & 0 < z < r_{cage} \\ \dfrac{(\lambda-1)z^3 - (\lambda-1)z^2 + 3(\lambda-1)\left(1-r_{cage}^2\right)z - (\lambda-1) + 3(\lambda-1)r_{cage}^2 + 2(\lambda+1)r_{cage}^3}{4r_{cage}^3} & r_{cage} < z < r_{cage}+d \end{cases} \tag{B4}$$

the local and collective elastic barriers at every position in the liquid can be computed

Figure B1 shows an example of how the local, elastic, and total barriers for a semi-infinite liquid vary as a function of distance from the surface for $\lambda = 1.038$. The increase in the local barrier is the primary reason that relaxation slows down near the interface. The elastic barrier is subject to competing tendencies near the interface, which leads to the unusual non-monotonic shape. The cutoff of the integral in Eq. B3 tends to depress the elastic barrier, but the liquid near the interface is stiffened compared to the bulk which tends to raise the barrier. The naive cutoff idea is, of course, not literally nor generically true for solid surfaces. But it does render no adjustable parameter calculations possible to perform by circumventing many real world issues which remain theoretically open such as the appropriate microscopic boundary conditions, the gradient in elasticity near the interface, etc.

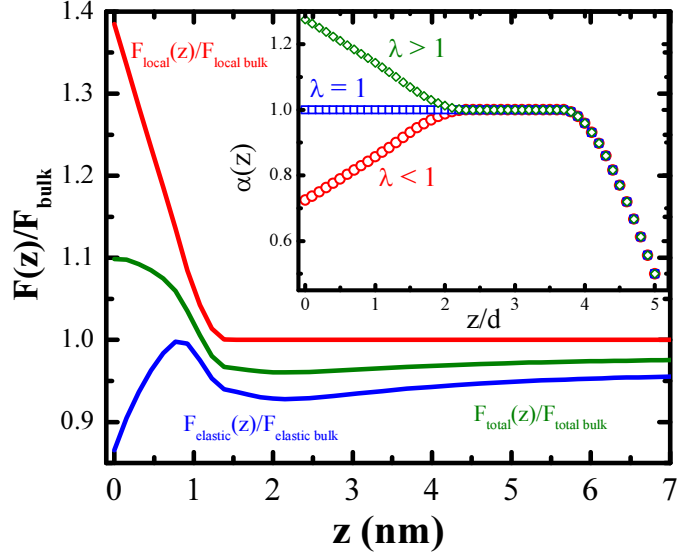

**Figure B1**: Normalized local (red line), elastic (blue line) and total barrier (green line) of the nanocomposites in terms of the distance $z$ away from the solid-liquid interface. The inset shows qualitative behavior of the $\alpha(z)$ under different $\lambda$ values.

The local shear modulus at a location $z$ in the film is approximately computed based on the standard mode-coupling-like formula [53]:

$$G(z) = \frac{k_B T}{60\pi^2} \int_0^\infty dk \left[ k^2 \frac{d}{dk} S(k;\varphi(z)) \right]^2 \exp\left( -\frac{k^2 r_{loc}^2(z)}{3 S(k;\varphi(z))} \right) \tag{B5}$$

where $r_{loc}(z)$ is the transient (harmonic) localization length computed from the dynamic free energy. To compute a layer-averaged, frequency-dependent elastic shear modulus we adopt a simple Maxwell model whence[51]:

$$\langle G'(z,\omega)\rangle = l_{layer}^{-1}\int_0^{l_{layer}} dz \frac{(\omega G(z))^2}{1+(\omega G(z))^2} \qquad (B6)$$

## Acknowledgements

We acknowledge the help of Dr. J. Borreguero Calvo for assistance in the glycerol force-field and Dr. H. Heinz for assistance in the amorphous silica force-field. We thank Dr. J. J. Towey and Dr. L. Dougan for providing us the neutron scattering data of glycerol. This work was supported by the U.S. Department of Energy, Office of Science, Basic Energy Sciences, Materials Sciences and Engineering Division. This research used resources of the Oak Ridge Leadership Computing Facility (OLCF) at the Oak Ridge National Laboratory, which is supported by the Office of Science of the U.S. Department of Energy under Contract No. DE-AC05-00OR22725.